\documentclass[aps,prl,twocolumn,groupedaddress]{revtex4-1}

\usepackage{color}
\usepackage{graphicx}
\usepackage{dcolumn}
\usepackage{bm}
\usepackage{float}
\usepackage[mathlines]{lineno}
\usepackage{tabularx}
\usepackage{hyperref}
\usepackage{footnote}
\usepackage{amsmath}
\usepackage{txfonts}

\begin{document}

\title{Phonon-mediated high-temperature superconductivity in ternary borohydride KB$_2$H$_8$ around 12 GPa}

\author{Miao Gao}\email{gaomiao@nbu.edu.cn}

\author{Xun-Wang Yan$^{2}$}
\author{Zhong-Yi Lu$^{3}$}
\author{Tao Xiang$^{4,5,6}$}

\date{\today}

\affiliation{$^1$Department of Physics, School of Physical Science and Technology, Ningbo University, Zhejiang 315211, China}

\affiliation{$^{2}$College of Physics and Engineering, Qufu Normal University, Shandong 273165, China}

\affiliation{$^{3}$Department of Physics, Renmin University of China, Beijing 100872, China}

\affiliation{$^{4}$Institute of Physics, Chinese Academy of Sciences, Beijing 100190, China }

\affiliation{$^{5}$School of Physical Sciences, University of Chinese Academy of Sciences, Beijing 100049, China}

\affiliation{$^{6}$Beijing Academy of Quantum Information Sciences, Beijing 100193, China}

\begin{abstract}
Discovery of high-temperature superconductivity in hydrogen-rich compounds has
fuelled the enthusiasm for finding materials with more promising superconducting properties among hydrides.
However, the ultrahigh pressure needed to synthesize and maintain high-temperature hydrogen-rich superconductors
hinders the experimental investigation of these materials.
For practical applications, it is also highly desired to find more hydrogen-rich materials that superconduct at high temperatures but under relatively lower pressures. Based on first-principles density functional theory, we calculate the electronic and phonon band structures for a ternary borohydride formed by intercalating BH$_4$ tetrahedrons into a face-centered-cubic potassium lattice, KB$_2$H$_8$. Remarkably, we find that this material is dynamically stable and one of its $sp^3$-hybridized $\sigma$-bonding bands is metallized (i.e. partially filled) above a moderate high pressure. This metallized $\sigma$-bonding band couples strongly with phonons, giving rise to a strong superconducting pairing potential. By solving the anisotropic Eliashberg equations, we predict that the superconducting transition temperature of this compound is 134-146 K around 12 GPa.
\end{abstract}

\maketitle

The road to room-temperature superconductivity traces back to the prediction that hydrogen, if sufficiently squeezed, could turn into a metal that superconducts at high temperature \cite{Ashcroft-PRL21,McMahon-PRB84}.
However, as the pressure for hydrogen metallization is exceptionally high and difficult to achieve \cite{Dubrovinsky-NC3,Dias-Science355}, an alternative route to high-temperature superconductivity was suggested, through chemical precompression in hydride materials \cite{Ashcroft-PRL92}. In 2015, hydrogen sulfide H$_3$S was found to superconduct at 203 K under 155 GPa \cite{Drozdov-Nature525}. This has rekindled the hope to realize near-room-temperature superconductivity through the phonon-mediated mechanism. Based on a particle swarm optimization algorithm, the crystal structures and superconducting properties in cage-like hydrogen-rich compounds, such as LaH$_{10}$, YH$_9$, and YH$_{10}$, under high pressure were systematically investigated \cite{Liu-PNAS114,Peng-PRL119}. The superconducting transition temperatures (T$_c$) were predicted to be 274-286 K for LaH$_{10}$ at 210 GPa, 253-276 K for YH$_9$ at 150 GPa, and 305-326 K for YH$_{10}$ at 250 GPa.
High-temperature superconductivity has been observed in LaH$_{10}$ and YH$_9$ experimentally \cite{Drozdov-Nature569,Somayazulu-PRL122,Snider-PRL126}.
Recently, a superconducting transition was reported at 288 K in a carbonaceous sulfur hydride at 267 GPa \cite{Snider-Nature586}.

While significant progress has been achieved in unveiling hydrogen-rich superconductors, physical characterization of these materials remain challenging experimentally. In particular, a diamond anvil cell (DAC) technique with a culet of 40-80 $\mu$m has to be utilized to produce a pressure above megabar \cite{Drozdov-Nature525,Drozdov-Nature569,Snider-Nature586}.
Such a small culet makes the electrode laying extremely difficult in the transport measurement.
Due to parochial sample chambers, the diameters of hydrogen-rich compounds were limited to 25-100 $\mu$m for H$_3$S \cite{Drozdov-Nature525} and 10-20 $\mu$m for LaH$_{10}$ \cite{Drozdov-Nature569} whose diamagnetic signal is almost undetectable.
Moreover, as the laser heating technology combined with DAC is used to modulate temperature during synthesis, it is difficult to control the purity and homogeneity of a sample due to large temperature gradient induced by laser.
In contrast, a multianvil large-volume press (LVP) could be used to generate a modest high pressure with uniform and well-controlled temperature in a larger sample chamber.
For example, the Kawai-type multianvil apparatus can generate
pressures to about 40 GPa with centimeter-sized sample volume \cite{Kawai-RSI41,Ishii-RSI87,Irifune-PEPI228}.
Thus, if the stable pressure of hydrogen-rich superconductors can be reduced to the working range of LVP, conventional physical measurements will be readily actualized.
On the theoretical side, current investigation focuses more on binary compounds, such as H$_3$S \cite{Duan-SR4}, LaH$_{10}$ \cite{Liu-PNAS114,Peng-PRL119}, Si$_2$H$_6$ \cite{Jin-PNAS107}, and CaH$_6$ \cite{Wang-PNAS109}. Ternary compounds are rarely explored, partly due to the computation complexity.

Superconducting transition temperatures of phonon-mediated superconductors are predominantly determined by two factors: the strength of electron-phonon coupling (EPC) and the characteristic energy scale of phonon excitations \cite{Bardeen-PR108,Allen-PRB12}. A universal strategy to enhance EPC with strong superconducting instability, as pointed out in Refs. \cite{Gao2015,Gao2020,Gao-Phys}, is to metalize one or a few electronic bands formed by the $\sigma$-bonding or other strong chemical bonding electrons. $\sigma$-bonds are the strongest type of covalent interactions and are formed through the head-to-head overlap of two atomic orbitals along the axial direction linking two atoms.
The $\sigma$-bonding bands are usually completely filled or unoccupied, without contribution to conductivity. However, its coupling with the vibrations of constituent atoms is very strong, which is in fact the main interaction that stabilizes the lattice structure. If a $\sigma$-band is lifted to the Fermi level and becomes conducting without distablizing the crystal structure, its coupling with phonons is partially released. However, the remanent EPC of this band could still be very strong, which may serve as a superglue to pair electrons with a large superconducting energy gap. A typical high-temperature superconducting compound that demonstrates this picture is MgB$_2$, which is a 39 K superconductor at ambient pressure, resulting from a strong coupling of a partially occupied $\sigma$-bonding band with phonons \cite{An-PRL86,Kong-PRB64,Kortus-PRL86}.

Taking a detailed analysis of the band structures with their EPC for hydrogen-rich superconducting materials whose first-principles results are available, we find that the metallization of certain hydrogen-related $\sigma$-bonding bands indeed plays a vital role in the formation of high-T$_c$ superconductivity \cite{Bernstein-PRB91,Peng-PRL119,Quan-PRB93,Liu-PRB99}. Furthermore, in order to take full advantage of the characteristic high energy scale of phonon modes of hydrogen atoms, a large contribution to the density of states (DOS) around the Fermi level by metallized hydrogen orbitals is also crucial. Thus there are two guiding rules that can be used for hunting high-$T_c$ superconductivity in hydrogen-rich compounds under high pressure: (1) existence of metallized hydrogen-related $\sigma$-bonding bands, and (2) a large fraction of density of states contributed by hydrogen orbitals around the Fermi level. For H$_3$S and other superhydride high-T$_c$ superconductors discovered so far under ultrahigh pressure, these two guiding rules are all satisfied \cite{Bernstein-PRB91,Peng-PRL119,Duan-SR4,Liu-PNAS114,Quan-PRB93,Liu-PRB99}.

However, the electronegativity of hydrogen is small in comparison with sulfur or other atoms with relatively large atomic numbers near ambient pressure. As a result, the contribution of H-$1s$ orbital to a metallized H-S or other $\sigma$-bonding state is relatively small even at ultrahigh pressure. The H-H bonding tends to combine two hydrogen atoms to a H$_2$ molecular, whose appearance is even adverse for high-$T_c$ superconductivity \cite{Fu-CM28,Zurek-PNAS106,Lonie-PRB87,Chen-NC12}.
Thus to find a hydrogen-rich high-T$_c$ superconductor at lower pressures, a non-metallic element with relatively small electronegativity and light atomic mass, such as boron, would be an ideal bonding partner of hydrogen.

\begin{figure}[t]
\begin{center}
\includegraphics[width=8.6cm]{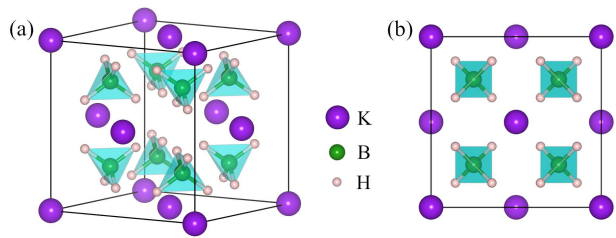}
\caption{Crystal structure of KB$_2$H$_8$. (a) Three-dimensional view. (b) View along [100] direction. Purple, green, and pink balls represent potassium, boron, and hydrogen atoms, respectively. The solid black line denotes the unit cell.}
\label{fig:Stru}
\end{center}
\end{figure}

In this work, we propose to use BH$_4$ tetrahedron as a building block to construct novel superconducting hydrides. BH$_4$ has a relatively high hydrogen content. It is also a building block of metal borohydride, such as Zr(BH$_4$)$_4$ \cite{Rude-JPCC116}. To obtain a stable and metallic compound, we intercalate a face-centered-cubic (fcc) lattice of potassium with BH$_4$. This yields the compound KB$_2$H$_8$ whose crystal structure is depicted in Fig.~\ref{fig:Stru}.
We calculate the electronic band structures, lattice dynamics, and the corresponding EPC for KB$_2$H$_8$ using first-principles density functional theory in combination with the Wannier interpolation method. A key feature revealed is that KB$_2$H$_8$ is dynamically stable in a pressure (12 GPa) that is significantly lower than that needed for stablizing the hydrogen sulfide superconductor H$_3$S. More importantly, we find that the two guiding rules above mentioned are perfectly fulfilled in KB$_2$H$_8$: one of the $sp^3$-hybridized $\sigma$-bonding bands is partially occupied, hence metallized, and the H-$1s$ orbital contributes about 41.0\% of the density of states at the Fermi level. Setting the Coulomb pseudopotential $\mu^*$ to a value between 0.10 and 0.15, we find its superconducting transition temperature T$_c\sim$ 134-146 K, by solving self-consistently the anisotropic Eliashberg equations.

In our calculations, the first-principles package, Quantum-ESPRESSO, is adopted \cite{pwscf}. We calculate the electronic states and the phonon perturbation potentials \cite{Giustino-PRB76} using the generalized gradient approximated Perdew-Burke-Ernzerhoff formula \cite{Perdew-PRL77}. The optimized norm-conserving Vanderbilt pseudopotentials are used to model the electron-ion interactions \cite{ONCV-PP}. After convergence test, the kinetic energy
cut-off and the charge density cut-off are set to 120 Ry and 480 Ry,
respectively. The charge densities are calculated on an unshifted 18$\times $18$\times $18 wave vector $\mathbf{k}$ mesh in combination with a Methfessel-Paxton smearing \cite{Methfessel-PRB40} of 0.02 Ry. The dynamical matrices and the perturbation potentials are computed on a $\Gamma $-centered 6$\times $6$\times $6 mesh, within
the framework of density-functional perturbation theory \cite{Baroni-RMP73_515}.
Eight $sp^3$-hybridized orbitals localized in the middle of boron-hydrogen bonds are used to construct the maximally localized Wannier functions (MLWFs) \cite{Pizzi-JPCM32}. The EPC constant $\lambda$ is evaluated with a fine electron (72$\times $72$\times $72) grid and a fine phonon (24$\times $24$\times $24) grid using the Electron-Phonon Wannier (EPW) package \cite{Ponce-CPC209}. The Dirac $\delta $-functions for electrons and phonons are smeared out by  Gaussian functions with the widths of 50 meV and 0.5 meV, respectively. Furthermore, a fine electron grid of 48$\times $48$\times $48 is employed in solving the anisotropic Eliashberg equations \cite{Ponce-CPC209,Choi-PC385,Margine-PRB87}. The Matsubara frequency is truncated at $\omega_c$=2.4 eV, which is about eight times of the highest phonon excitation energy.

Figure \ref{fig:Stru} shows the crystal structure of KB$_2$H$_8$. Each BH$_4$ tetrahedron is surrounded by four potassium atoms, which also form a tetrahedron.
The space group of KB$_2$H$_8$ is $Fm\bar{3}m$ (No. 225). The potassium, boron, and hydrogen atoms occupy the $4a$ (0.000, 0.000, 0.000), $8c$ (0.250, 0.250, 0.250), and $32f$ (0.357, 0.357, 0.357) Wyckoff positions, respectively.
This structure is stable above 12 GPa, close to the critical pressure (11.5 GPa) for the pure potassium crystal transforming from the body-centered cubic to the fcc phase \cite{Takemura-PRB28,Winzenick-PRB50}. The B-H bond length is about 1.2272 {\AA} at 12 GPa, shorter than the H-S bond of 1.5461 {\AA} in H$_3$S at 150 GPa \cite{Einaga-NP12}. This suggests that the covalent B-H bond is very strong in this material. Under ambient pressure, if the lattice instability is not considered, the optimized lattice constant of KB$_2$H$_8$ is about 6.6213 {\AA}, which is exactly the lattice constant of the fcc potassium under the ambient pressure.

\begin{figure}[t]
\begin{center}
\includegraphics[width=8.6cm]{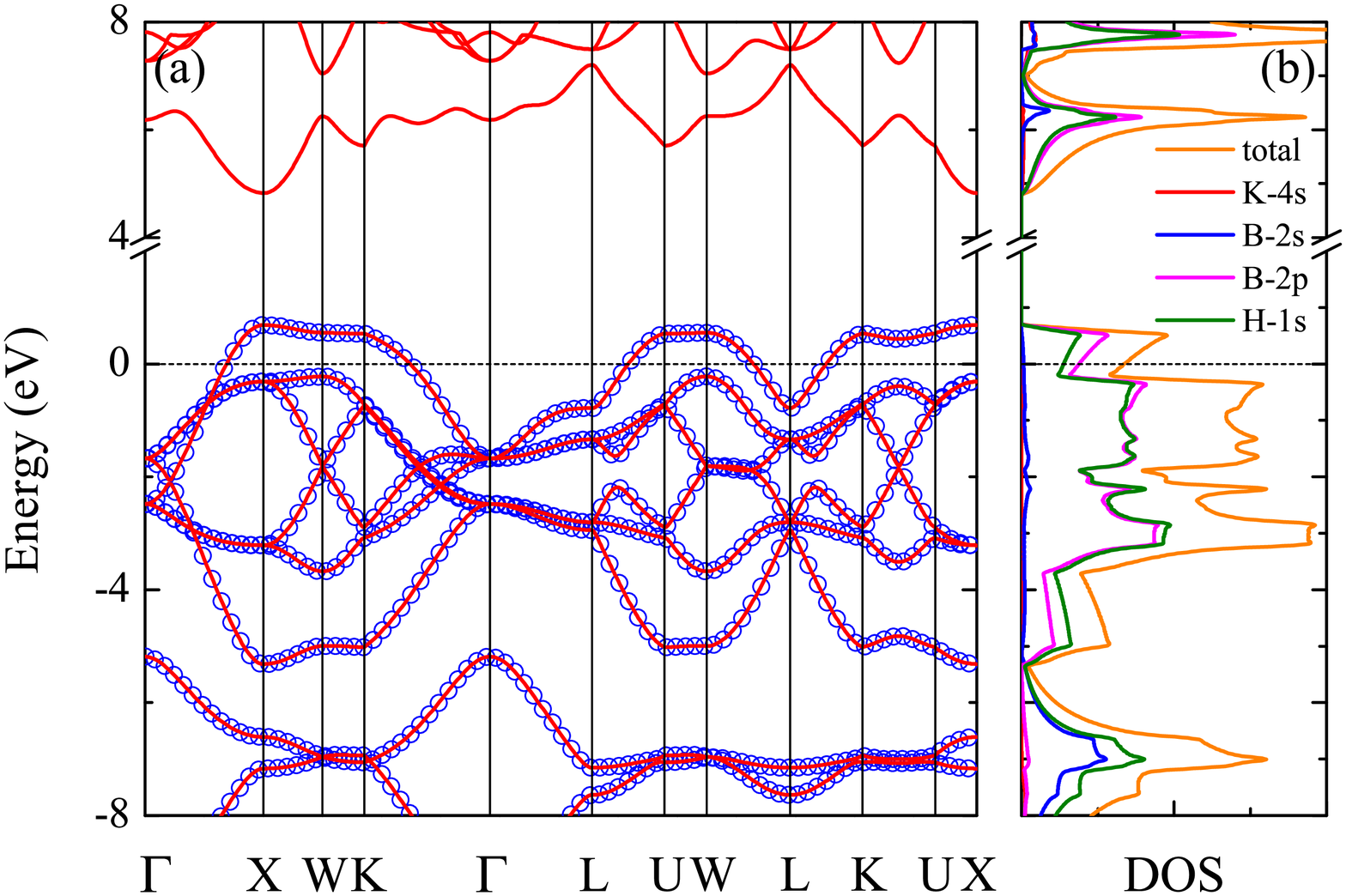}
\includegraphics[width=8.6cm]{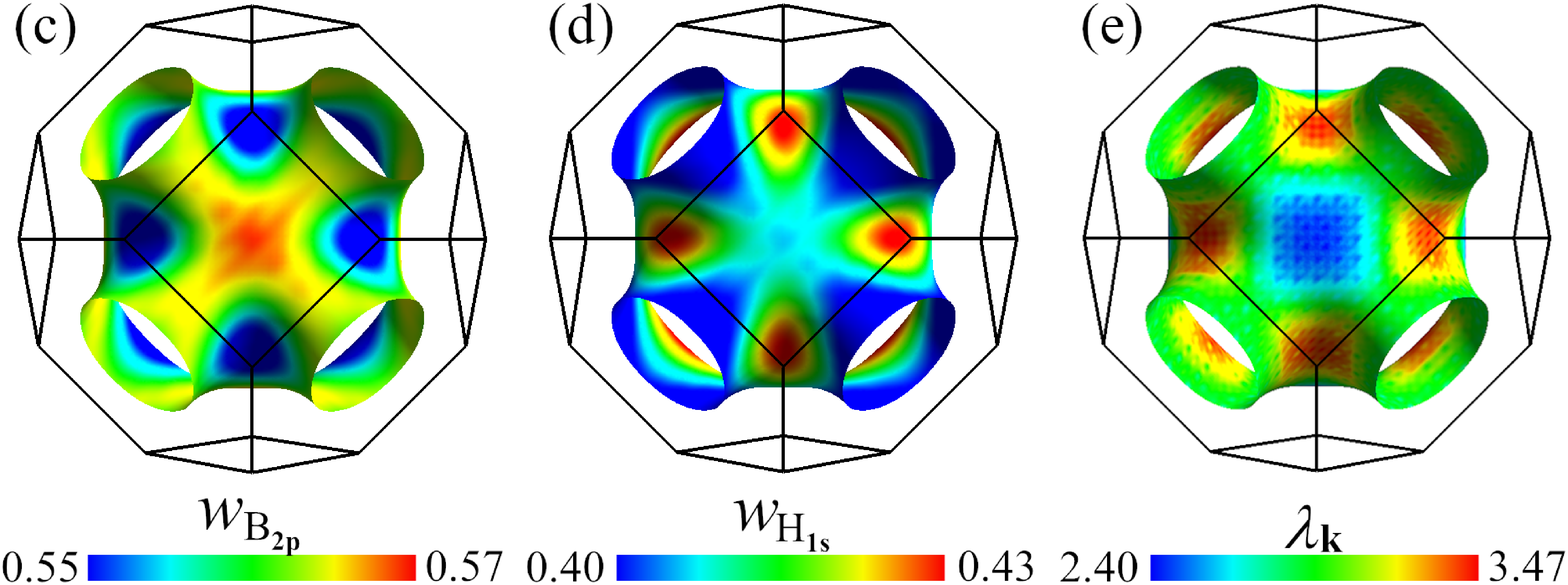}
\caption{Electronic structure of KB$_2$H$_8$. (a) Band structure: The red lines and blue circles denote the band structures obtained by the first-principles calculation and the MLWFs interpolation, respectively. The Fermi level is set to zero. (b) Total and orbital-resolved density of states. (c-d) Spectral weight on the Fermi surfaces for (c) the B-$2p$ orbitals and (d) the H-$1s$ orbitals, respectively. (e) The strength of EPC $\lambda_{{\bf k}}$ on the Fermi surface.}
\label{fig:Band}
\end{center}
\end{figure}

Figure \ref{fig:Band} shows the electronic structure of KB$_2$H$_8$ at 12 GPa.
There is a direct energy gap about 4.13 eV at the $X$ point [Fig.~\ref{fig:Band}(a)].
The consistency between first-principles band and the one generated through MLWFs interpolation shows
unambiguously that $sp^3$-hybridized $\sigma$ bonds are formed between boron and hydrogen atoms.
For the energy scale investigated, boron and hydrogen orbitals have almost the same contribution to DOS [Fig.~\ref{fig:Band}(b)],
due to their adjacent electronegativity.
There is only one band crossing the Fermi level.
The necks of the Fermi surface are truncated at the hexagonal boundary of the Brillouin zone.
By mapping the spectral weights of B-$2p$ and H-$1s$ orbitals and the wave vector {\bf k}-resolved EPC constant $\lambda_{{\bf k}}$
onto the Fermi surface, we find that the distribution of $\lambda_{{\bf k}}$ is similar to the spectral weight of H-$1s$, especially at the saddle points that link two necks.
This indicates that the H-$1s$-related electronic states couple strongly with phonons.

\begin{figure}[tbh]
\begin{center}
\includegraphics[width=8.6cm]{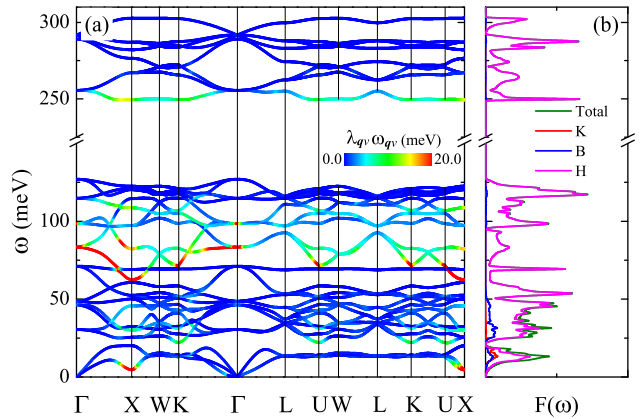}
\caption{Lattice dynamics of KB$_2$H$_8$. (a) Phonon spectrum with a false-color representation of $\lambda_{{\bf q}v}$$\omega_{{\bf q}v}$.
(b) Projected phonon density of states generated using the quasi-harmonic approximation. }
\label{fig:phonon}
\end{center}
\end{figure}

The lattice structure of KB$_2$H$_8$ is dynamically stable from 12 GPa up to at least 200 GPa. Figure \ref{fig:phonon} shows the phonon spectrum of KB$_2$H$_8$ at 12 GPa. As an indication of lattice stability, no imaginary phonon modes appear in the spectral function in this pressure range. The phonon spectrum is separated into two parts with an energy gap of 122.43 meV [Fig.~\ref{fig:phonon}(a)]. As expected, the phonon density of states $F(\omega )$ is dominated by hydrogen-associated phonon modes in the whole energy range shown in  Fig.~\ref{fig:phonon}(b). In particular, the high-frequency phonon excitations above 50 meV contribute predominantly by hydrogen atoms. On the contrary, the phonon excitations contributed by potassium and boron atoms exist mainly below 50 meV. The hydrogen phonons around 80 meV couple strongly with electrons. Several acoustic modes, for example, the acoustic phonon modes around the $X$ point, also have sizeable EPC. From the phonon displacement calculation, we find that the strong EPC in the acoustic branch around $X$ also results mainly from the vibration of hydrogen atoms. Hence, the hydrogen-associated phonon modes play a regnant role in coupling with electrons through H-$1s$ orbitals.

\begin{figure}[t]
\begin{center}
\includegraphics[width=8.6cm]{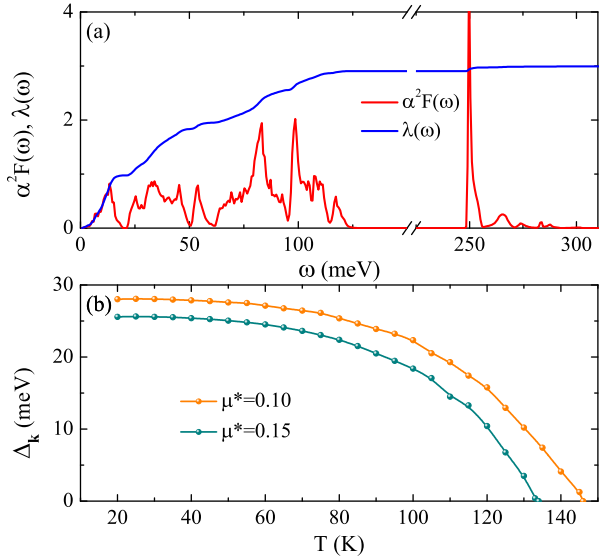}
\caption{(a) Eliashberg spectral function $\alpha^2F(\omega)$ and the accumulated EPC $\lambda(\omega)$, and
(b) temperature dependence of the average energy gap $\Delta_{\bf k}$, for KB$_2$H$_8$ at 12 GPa.}
\label{fig:a2f}
\end{center}
\end{figure}

We calculate the isotropic Eliashberg spectral function $\alpha^2F(\omega)$ through the MLWFs interpolation approach at 12 GPa. The result, together with the accumulated EPC
\begin{equation}
\lambda(\omega) = 2\int_0^\omega \frac{1}{\omega'} \alpha^2F(\omega') d\omega' ,
\end{equation}
is shown in Fig.~\ref{fig:a2f}(a). The full EPC parameter in the $\omega\rightarrow \infty$, $\lambda$, and the logarithmic average frequency, $\omega _{\text{log}}$, are equal to 2.99 and 32.84 meV, respectively. The value of this $\lambda$ is about 36\% higher than the corresponding values for the high-temperature superconductors H$_3$S and LaH$_{10}$ in the ultrahigh pressures. From the frequency dependence of $\lambda(\omega)$, we know that about one third of the total $\lambda$ comes from the contribution of acoustic phonons.
$\alpha^2F(\omega)$ shows a sharp peak around 250 meV. However, the high-energy phonon excitations above the gap contribute just 2.85\% of the total $\lambda$. This is the reason why the value of $\omega _{\text{log}}$ in KB$_2$H$_8$ is much lower than the corresponding value in H$_3$S \cite{Duan-SR4}.

While only one band crosses the Fermi level in KB$_2$H$_8$  [Fig.~\ref{fig:Band}(a)], the distribution of EPC on the Fermi surface exhibits visible anisotropy [Fig.~\ref{fig:Band}(e)]. To determine the superconducting transition temperature for this material, we solve the anisotropic Eliashberg equations. By setting the Coulomb pseudopotential $\mu^*$ to 0.10-0.15, we find that the superconducting transition temperature is about 134-146 K at 12 GPa [Fig.~\ref{fig:a2f}(b)]. At 20 K, the average energy gap $\Delta_{\bf k}$ is about 28.03 meV. There is a hardening in the phonon excitation modes of KB$_2$H$_8$ at a higher pressure. The characteristic frequency $\omega_{\text{log}}$ becomes 57.37 meV at 20 GPa. The corresponding EPC constant $\lambda$, however, drops to 1.69, which reduces the superconducting transiton temperature to 113-123 K at that pressure.

We have also calculated the formation enthalpy of KB$_2$H$_8$ under pressure. Our calculation shows that the formation enthalpy of KB$_2$H$_8$ is -1.76 eV/formula at 12 GPa and -2.29 eV/formula at 20 GPa, in comparison with the fcc potassium, $\alpha$-B$_{12}$ \cite{Oganov-Nature457}, and $P6_3/m$ hydrogen \cite{Pickard-NP3}. This suggests that there is a high probability to synthesize successfully KB$_2$H$_8$ at 12 GPa or above.

In summary, we have proposed two guiding rules, based on the physical picture that the EPC can be significantly enhanced if the $\sigma$-bonding bands can be metallized without distablizing the lattice structure, for hunting phonon-mediated high-T$_c$ superconductors in superhydrides. We take KB$_2$H$_8$ as an example and carry out a first-principles investigation on its electronic band structure and lattice dynamics. It is shown that this compound is metallized and stable when the pressure is above 12 GPa. More importantly, we find that coupling between H-$1s$ orbital and H-associated phonons is very strong. By solving the anisotropic Eliashberg equations, the superconducting transition temperature of this superhydride is found to be above 134 K at 12 GPa and above 113 K at 20 GPa.

This work was supported by the National Natural Science Foundation of China (Grant Nos. 11974194, 11974207, 11888101, and 11934020) and the National Key Research and Development Program of China (Grant No. 2017YFA0302900).
M.G. was sponsored by K. C. Wong Magna Fund in Ningbo University.

\end{document}